\documentclass[conference]{IEEEtran}
\IEEEoverridecommandlockouts

\usepackage{hyperref} 
\usepackage{xcolor}
\usepackage{fancybox}
\usepackage{fancyhdr}

\usepackage{graphicx} 
\usepackage{makecell}
\usepackage{listings} 
\usepackage{multirow}
\usepackage{xurl}

\usepackage{url}
\usepackage{booktabs}

\usepackage{cite}
\usepackage{amsmath,amssymb,amsfonts}
\usepackage{algorithmic}
\usepackage{textcomp}
\usepackage[normalem]{ulem} 
\usepackage{placeins}
\usepackage{balance} 

\newcommand{\hypobox}[1]{%
    \begin{center}%
        \noindent\thicklines\setlength{\fboxsep}{2pt}%
        \cornersize{0.1}%
        \ovalbox{%
            \begin{minipage}{0.90\linewidth}%
                #1%
            \end{minipage}%
        }%
    \end{center}%
}

\newcommand*{\RQTwo} [1] {And isn't this one even better ?}
\newcommand*{\RQThree} [1] {This one tops it all, doesn't it ?}

\definecolor{keywordcolor}{rgb}{0.0, 0.2, 0.6} 
\definecolor{commentcolor}{rgb}{0.3, 0.6, 0.3} 
\definecolor{stringcolor}{rgb}{0.2, 0.5, 0.2} 
\definecolor{classcolor}{rgb}{0.7, 0.5, 0.0} 
\definecolor{annotationcolor}{rgb}{0.6, 0.3, 0.0} 
\definecolor{numbercolor}{rgb}{0.4, 0.4, 0.4} 

\lstdefinestyle{javacode}{
  language=Java,
  basicstyle=\footnotesize\color{black}, 
  keywordstyle=\color{keywordcolor}, 
  commentstyle=\color{brown}\itshape, 
  stringstyle=\color{stringcolor}, 
  classoffset=1, 
  morekeywords={UploadImage}, 
  keywordstyle=[1]\color{classcolor}, 
  keywordstyle=[2]\color{annotationcolor}, 
  morekeywords=[2]{@Test}, 
  numbers=left,
  numberstyle=\tiny\color{numbercolor}, 
  stepnumber=1,
  frame=single,
  tabsize=4,
  showstringspaces=false,
  breaklines=true,
  captionpos=b,
  numbersep=10pt, 
  framexleftmargin=15pt, 
  xleftmargin=20pt 
}

\lstdefinestyle{pythoncode}{
  language=Python,
  basicstyle=\ttfamily\footnotesize\color{black}, 
  keywordstyle=\color{keywordcolor}, 
  commentstyle=\color{commentcolor}\itshape, 
  stringstyle=\color{stringcolor}, 
  numbers=left,
  numberstyle=\tiny\color{numbercolor}, 
  stepnumber=1,
  frame=single,
  tabsize=4,
  showstringspaces=false,
  breaklines=true,
  captionpos=b,
  numbersep=10pt, 
  framexleftmargin=15pt, 
  xleftmargin=20pt, 
  morekeywords={self,True,False,None,async,await,with,as,lambda}, 
  emph={print,len,range,open,str,int,float,list,dict,set,tuple,enumerate,zip,Path},
  emphstyle=\color{classcolor}
}

\lstdefinestyle{pythoncommand}{
  basicstyle=\ttfamily\footnotesize\color{black},
  keywordstyle=\color{keywordcolor},
  commentstyle=\color{commentcolor}\itshape,
  stringstyle=\color{stringcolor},
  numbers=none,
  frame=single,
  tabsize=2,
  showstringspaces=false,
  breaklines=true,
  breakatwhitespace=true,
  keepspaces=true,
  columns=fullflexible,
  captionpos=b,
  xleftmargin=10pt,
  framexleftmargin=8pt,
  xrightmargin=5pt,
  morekeywords={python,restcov.py},
  literate=
    {--openapi}{{{\color{classcolor}--openapi}}}9
    {--logs}{{{\color{classcolor}--logs}}}6
    {--out}{{{\color{classcolor}--out}}}5
}

\lstdefinelanguage{jsonlang}{
    basicstyle=\ttfamily\footnotesize,
    showstringspaces=false,
    breaklines=true,
    morestring=[b]",
    stringstyle=\color{stringcolor},
    literate=
     *{0}{{{\color{numbercolor}0}}}{1}
      {1}{{{\color{numbercolor}1}}}{1}
      {2}{{{\color{numbercolor}2}}}{1}
      {3}{{{\color{numbercolor}3}}}{1}
      {4}{{{\color{numbercolor}4}}}{1}
      {5}{{{\color{numbercolor}5}}}{1}
      {6}{{{\color{numbercolor}6}}}{1}
      {7}{{{\color{numbercolor}7}}}{1}
      {8}{{{\color{numbercolor}8}}}{1}
      {9}{{{\color{numbercolor}9}}}{1}
      {:}{{{\color{black}:}}}{1}
      {,}{{{\color{black},}}}{1}
      {\{}{{{\color{keywordcolor}\{}}}{1}
      {\}}{{{\color{keywordcolor}\}}}}{1}
      {[}{{{\color{keywordcolor}[}}}{1}
      {]}{{{\color{keywordcolor}]}}}{1},
}

\lstdefinestyle{json}{
    language=jsonlang,
    basicstyle=\ttfamily\footnotesize\color{black},
    numbers=left,
    numberstyle=\tiny\color{numbercolor},
    stepnumber=1,
    numbersep=8pt,
    frame=single,
    tabsize=2,
    showstringspaces=false,
    breaklines=true,
    breakatwhitespace=true,
    keepspaces=true,
    columns=fullflexible,
    captionpos=b,
    xleftmargin=14pt,
    framexleftmargin=10pt,
    xrightmargin=5pt
}

\hypersetup{
    colorlinks,%
    citecolor=black,%
    filecolor=black,%
    linkcolor=black,%
    urlcolor=gray,
    pdftitle={The Next level of Software Test Automation},    
    pdfauthor={Sigrid Eldh Serge Demeyer},     
    pdfsubject={Book},   
    pdfcreator={PDF LaTex},   
}
\def\BibTeX{{\rm B\kern-.05em{\sc i\kern-.025em b}\kern-.08em
    T\kern-.1667em\lower.7ex\hbox{E}\kern-.125emX}}
\begin{document}

\title{RESTCov: A Tool for Structural Coverage Analysis of REST APIs\\
\thanks{This work is supported by the Research Foundation Flanders (FWO) via the BaseCamp Zero Project under Grant number S000323N.}
}

\author{
\IEEEauthorblockN{Tolgahan Bardakci}
\IEEEauthorblockA{Computer Science \\
University of Antwerp and Flanders Make \\
Antwerp, Belgium \\
tolgahan.bardakci@uantwerpen.be}
\and
\IEEEauthorblockN{Serge Demeyer}
\IEEEauthorblockA{Computer Science \\
University of Antwerp and Flanders Make \\
Antwerp, Belgium \\
serge.demeyer@uantwerpen.be}
}

\maketitle

\begin{abstract}
REST APIs are widely used in modern software systems, but developers and testers often lack visibility into which parts of an API specification are exercised by a test suite.
Traditional coverage analysis usually relies on source-code instrumentation, which is impractical for REST APIs that are distributed, externally maintained, and hence accessible only through black-box execution.
This paper presents RESTCov, a lightweight tool that computes structural REST API coverage from an OpenAPI specification and observed HTTP request/response logs, reporting coverage across paths, operations, parameters, media types, status codes, and status classes.
RESTCov produces both machine-readable results and a human-readable HTML report,
helping users inspect coverage gaps, diagnose specification-log mismatches, and evaluate REST API test suites without requiring access to the implementation.

Screencast:~\url{https://youtu.be/mNz2P43OyUc}

Repository:~\url{https://github.com/2tolgahan2/RESTCov}
\end{abstract}


\section{Introduction}
REST APIs are central interfaces for modern software systems, enabling communication between services, applications, and organizations~\cite{fielding2000architecturalstylesdesign, verborgh2015fallacymultiapiculture}.
As these APIs evolve, their OpenAPI specification~\cite{openapiinitiative2025openapispecification} and their actual implementation can drift out of sync: endpoints can be added, removed, or change behavior without a corresponding update to the specification.
Testing plays an important role in surfacing this drift by verifying that the documented behavior is exercised by existing test suites or observed in client interactions.
For developers and testers, this requires visibility into which parts of an API have actually been exercised.

Coverage analysis is a widely used method for assessing test sufficiency and understanding which parts of a system have been exercised~\cite{zhu1997softwareunittest}.
Coverage is often computed through source-code instrumentation.
However, for REST APIs, this is not always possible or desirable.
The implementation may be unavailable, distributed across multiple services, or maintained by another team.
In these situations, testers may still have access to two artifacts: an API specification and logs of observed HTTP traffic.
Together, these artifacts enable black-box coverage analysis by comparing executed requests with the documented API behavior.

OpenAPI specifications~\cite{openapiinitiative2025openapispecification} describe structural elements of REST APIs, such as paths, operations, parameters, media types, and expected response codes.
These elements provide a practical basis for measuring how much of a documented API has been exercised.
However, in practice, mapping observed traffic back to a specification is not always straightforward.
Logs may contain requests that cannot be matched to any documented operation in the OpenAPI specification.
We refer to these as~\emph{unmatched requests}.
This may happen because the request targets an undocumented endpoint, the path is malformed or encoded differently, or the implementation and specification have evolved out of sync~\cite{kim2022automatedtestgeneration}.
If such cases are hidden, coverage results become harder to interpret.
The missing coverage may reflect either untested API behavior or traffic that falls outside the documented API.

To address this need, we present RESTCov, a lightweight tool for black-box coverage analysis of REST APIs.
RESTCov takes an OpenAPI specification and observed HTTP logs, maps executed requests to documented API elements, and reports coverage across paths, operations, parameters, media types, status codes, and status classes.
It produces machine-readable results and a human-readable HTML report to inspect coverage gaps and diagnose unmatched requests.
This supports API maintenance and evolution by helping developers assess regression suites, detect specification-log drift, and prioritize test or documentation updates.

This paper makes the following contributions:
\begin{itemize}
    \item We present RESTCov, a lightweight tool for computing structural REST API coverage from an OpenAPI specification and observed HTTP traffic, without requiring source-code instrumentation.
    \item We describe the coverage dimensions supported by RESTCov, including paths, operations, parameters, media types, status codes, and status classes.
    \item We demonstrate RESTCov's machine- and human-readable outputs, including unmatched request samples to diagnose specification-log mismatches.
    \item We provide a reusable artifact with source code, example inputs, documentation, and expected outputs to support practical use and future experimentation.
\end{itemize}

\section{Related Work and Positioning}
\subsection{REST API Testing}
REST API testing has been widely studied~\cite{golmohammadi2023testingrestfulapis}. 
RESTTESTGEN~\cite{viglianisi2020resttestgenautomatedblackbox} and RESTest~\cite{martin-lopez2021restestautomatedblackbox} support automated black-box testing by generating test cases from API specifications.
Kim et al. further examine automated test generation for REST APIs and highlight the challenges of generating effective tests~\cite {kim2022automatedtestgeneration}.
In addition to test generation, Nooyens et al.~\cite{nooyens2026testamplificationrest} introduce agentic test amplification systems, based on single- and multi-agent LLMs, that extend existing REST API test suites.

Other tools extend REST API testing in different directions.
RESTler~\cite{atlidakis2019restlerstatefulrest} uses OpenAPI specifications for stateful REST API fuzzing, while EvoMaster~\cite{arcuri2018evomasterevolutionarymulticontext} targets automated system-level test generation for RESTful services.
QuickREST~\cite{karlsson2020quickrestpropertybasedtest} explores property-based test generation from OpenAPI specifications.
Schema-aware and model-based directions have also been explored through tools such as Schemathesis~\cite{hatfield-dodds2022derivingsemanticsawarefuzzers}, Morest~\cite{liu2022morestmodelbasedrestful}, and RestCT~\cite{wu2022combinatorialtestingrestful}.
These tools primarily focus on generating or executing REST API tests.

\subsection{REST API Coverage Analysis}
Coverage criteria for REST APIs have been proposed to measure how thoroughly tests exercise API specifications~\cite{martin-lopez2019testcoveragecriteria}.
Restats provides tool support for measuring REST API test coverage~\cite{corradini2021restatstestcoverage}.
RESTCov shares this motivation: making API-level coverage visible without requiring source-code access.

The two tools differ most in how they handle traffic that falls outside the documented specification.
Restats records some of this, for example through a~\emph{notDocumentedAndTested} section in its operation-coverage report.
However, that information is split across separate per-metric JSON files, with no single place to see it all.
RESTCov instead collects all such traffic into one explicit list of unmatched requests, shown directly in both its JSON output and its HTML report.
This makes it easy to spot specification-log mismatches, whether they stem from specification-implementation drift, such as undocumented endpoints, or from client-side artifacts, such as malformed or encoded paths, without digging through multiple reports.

\section{RESTCov Overview}
\subsection{Purpose}
\label{subsec:purpose}
RESTCov is designed for situations where users have an OpenAPI specification and observed HTTP traffic, but need to understand which parts of the documented API were exercised.
It also shows traffic that no longer corresponds to the specification, a sign that the implementation and specification may have drifted out of sync.
Rather than relying on source code, RESTCov measures coverage directly from execution logs and maps those observations back to the specification.

RESTCov reports whether a documented specification element (e.g., a path, operation, or status code) was observed in the traffic.
It does not verify that the corresponding functionality behaves correctly, nor does it assess the thoroughness or adequacy of the testing that produced this traffic.
As such, RESTCov measures coverage of the specification's structural elements, not functional correctness of the API's behavior.

\subsection{Inputs}
RESTCov operates on two inputs: an OpenAPI specification~\cite{openapiinitiative2025openapispecification}, and a directory of paired request/response log files.
It supports different OpenAPI versions, making it suitable for specifications generated by different tools or processes. 
In the current public version, each request is stored in a \texttt{*-request.txt} file and each response in a matching \texttt{*-response.txt} file.
RESTCov scans these and parses them for comparison against the documented API.

\subsection{Workflow}
RESTCov first parses the OpenAPI specification and extracts the documented paths, operations, parameters, response codes, and media types.
It then reads the request/response log pairs, and reconstructs the observed traffic, including the HTTP method, request path, visible parameters, request content type, response status code, and response content type.

After parsing, RESTCov matches each observed request to a documented API operation based on the request path and method.
Each documented path template (e.g., /pet/\{petId\}) is compiled into a regular expression that replaces path parameters with a wildcard segment, then matched against the request path after stripping any server or base-path prefix declared in the specification or supplied via the \texttt{--base-path} option.

Different encodings of the same path are common in REST APIs.
For example, a slash may appear literally or encoded as~\emph{\%2F}.
RESTCov requires an exact match, including casing, trailing slashes, and encoding.
A request that differs in any of these is reported as an unmatched request rather than matched to the corresponding operation.
In contrast, HTTP methods are compared case-insensitively.
Once a request is matched, RESTCov records which documented elements were observed in that interaction.
These observations are then aggregated into coverage metrics and written to the output files.

\subsection{Coverage Dimensions}
RESTCov reports coverage across several dimensions, comparing documented API elements against observed traffic, inspired by specification-based REST API coverage criteria~\cite{martin-lopez2019testcoveragecriteria}.

\begin{itemize}
    \item \textbf{Path coverage} measures whether a documented API path appears.
    \item \textbf{Operation coverage} refines path coverage by considering the combination of an HTTP method and a documented path, such as \texttt{GET /pet/\{petId\}} or \texttt{POST /pet}.
    \item \textbf{Parameter coverage} measures whether documented parameters are observed.
    \item \textbf{Request content-type coverage} measures whether the documented media types for request bodies are observed.
    \item \textbf{Response content-type coverage} measures whether the documented media types for response bodies are observed.
    \item \textbf{Status code coverage} measures whether documented response codes, such as \texttt{200}, \texttt{400}, or \texttt{404}, are observed.
    \item \textbf{Status class coverage} measures whether documented response codes are observed at the class level, such as \texttt{2xx}, \texttt{4xx}, or \texttt{5xx}, rather than as individual codes.
\end{itemize}

These dimensions give a diverse view of the system under test.
A test suite may exercise many operations while still missing parameters, media types, or response outcomes.
RESTCov makes such gaps visible by reporting each dimension.

\subsection{Outputs}
RESTCov produces two outputs: a machine-readable \texttt{coverage.json} file and a human-readable \texttt{report.html} file. 
The JSON file stores the computed metrics and observations for further analysis, test-suite comparison, or CI/CD integration. 
The HTML report presents the same information in an inspectable form and includes unmatched request samples.

The unmatched samples make specification-implementation drift directly visible.
Traffic that no longer corresponds to the specification may indicate that an endpoint was added, removed, or changed after the specification was last updated.
Not every mismatch reflects drift from the specification.
Some of the mismatches may come from client-side artifacts, such as malformed or encoded paths, rather than changes on the API side.
Once users know which of the two applies, they can update the specification if the implementation has moved on, or fix the client if the request itself is malformed.

\section{Demonstration}
\label{sec:demonstration}
This section illustrates how RESTCov analyzes black-box REST API coverage from an OpenAPI specification and observed HTTP traffic.
The demonstration uses the Petstore API~\cite{swaggerapi2026swaggerpetstoreapi}, a small open-source example, together with recorded request/response logs.
The goal is not to evaluate a specific test suite, but to show how RESTCov processes its inputs, produces coverage results, and helps users inspect gaps between documented and observed behavior.

While this demonstration uses a single API, RESTCov has also been used across five production APIs, including Google Drive and Spotify, in a comparative study~\cite{besjes2025agenticllmsrest} and an industrial setting~\cite{bardakci2026promptbasedrestapi}.
When applied to the OpenAPI specifications used across these studies, together with Petstore, RESTCov processed all of them successfully, while some of the other coverage tools we tested did not.
This underscores the importance of robustness across diverse, real-world OpenAPI documents.

\subsection{Command-Line Execution}
RESTCov runs from the command line by specifying the OpenAPI specification, log directory, and output directory:

\begin{lstlisting}[style=pythoncommand, caption={Example RESTCov command-line invocation}]
python restcov.py \
--openapi examples/petstore/openapi.json \
--logs examples/petstore/logs \
--out output/petstore
\end{lstlisting}

After execution, RESTCov parses the specification, reconstructs HTTP interactions from the logs, and, when possible, matches each request to a documented operation.
These observations are aggregated into coverage metrics and written to a machine-readable JSON file and a human-readable HTML report in the output directory.

Table~\ref{tab:petstore-coverage-results} summarizes coverage results.
For Path and Operation, the total counts are distinct documented elements; for other dimensions, it is an aggregate summed across all operations.

\begin{table}[htbp]
\centering
\caption{Coverage summary for the Petstore demonstration.}
\label{tab:petstore-coverage-results}
\begin{tabular}{lccc}
\hline
\textbf{Dimension} & \textbf{Total} & \textbf{Covered} & \textbf{Coverage} \\
\hline
Path & 14 & 14 & 100.0\% \\
Operation & 20 & 20 & 100.0\% \\
Parameter & 25 & 25 & 100.0\% \\
Request content type & 11 & 9 & 81.8\% \\
Response content type & 38 & 33 & 86.8\% \\
Status code & 36 & 14 & 38.9\% \\
Status class & 23 & 15 & 65.2\% \\
\hline
\end{tabular}
\end{table}

The observed traffic covers all documented paths, operations, and parameters.
However, coverage is lower for request content types, response content types, status codes, and status classes.
This indicates that reaching all documented operations does not guarantee that documented media types and response outcomes have been exercised.

\subsection{Output Interpretation}
Listing~\ref{lst:coverage-json} shows selected metrics and unmatched request samples from the generated \texttt{coverage.json} file.
The~\emph{total\_requests} field counts only the observed requests that were successfully matched to a documented operation; unmatched requests, such as the one shown in Listing~\ref{lst:coverage-json}, are excluded from this count and reported under \emph{unmatched\_samples} separately.

\begin{lstlisting}[style=json, caption={JSON report from the generated~\texttt{coverage.json} file.}, label={lst:coverage-json}]
{
  "spec": "openapi.json",
  "logs": "logs",
  "metrics": {
    "total_requests": 139,
    "total_paths": 14,
    "covered_paths": 14,
    "path_cov_pct": 100.0,
    "total_ops": 20,
    "covered_ops": 20,
    "op_cov_pct": 100.0,
    "status_total": 36,
    "status_seen": 14,
    "status_cov_pct": 38.9
  },
  "unmatched_samples": [
    {
      "method": "POST",
      "path": "/pet%2F%2FuploadImage"
    }
  ]
}
\end{lstlisting}

\begin{figure*}[!t]
    \centering
    \includegraphics[width=0.75\textwidth]{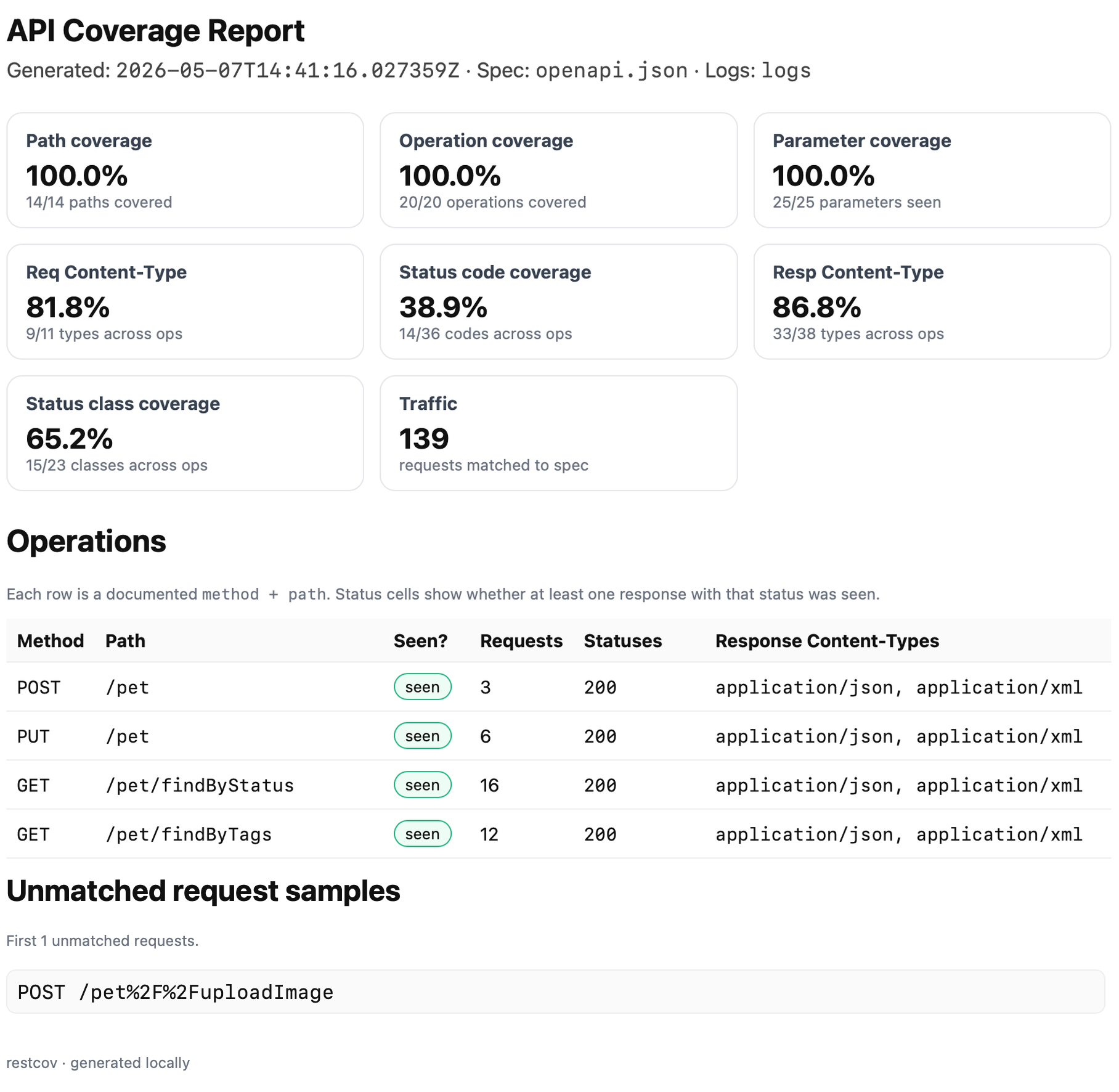}
    \caption{Excerpt from the generated \texttt{report.html} file for the Petstore demonstration.}
    \label{fig:restcov-html-report}
\end{figure*}

Figure~\ref{fig:restcov-html-report} shows an excerpt from the \texttt{report.html} file.
The report summarizes coverage by dimension and lists unmatched samples.
Such samples may correspond to undocumented endpoints, malformed or encoded paths, or other specification-log mismatches.
For example, the unmatched path~\textit{/pet\%2F\%2FuploadImage} contains a double-encoded slash (\texttt{\%2F\%2F}).
This is a client-side artifact, likely caused by concatenating an already encoded path segment, rather than a sign of an undocumented or uncovered operation.

\FloatBarrier

\subsection{Insights from the Report}
The report helps users move beyond a single coverage number.
For example, low parameter coverage may mean tests exercise only default or minimal requests.
In contrast, low status code coverage may mean tests mainly cover successful responses and miss error-handling behavior.
The unmatched samples also help users decide whether to add tests, improve the specification, or inspect the logging format.
Such mismatches often emerge as an API evolves: new endpoints go undocumented, client behavior changes, or specifications fall out of sync with implementation.
By surfacing these mismatches directly from traffic, RESTCov supports ongoing API maintenance and evolution, not just a one-time coverage check at release.

\hypobox{
The Petstore demonstration shows that RESTCov can reveal differences between exercised endpoints and exercised response behavior, and expose specification-log mismatches through unmatched request samples.
}

\section{Availability and Reuse}
RESTCov is available as an artifact on GitHub\footnote{\url{https://github.com/2tolgahan2/RESTCov}}.
The repository includes the source code, documentation, example inputs, expected outputs, and screenshots of the HTML report.
It is organized as follows:
\begin{itemize}
    \item The \texttt{root directory} contains the analysis script, \mbox{\textit{restcov.py}}, the dependency file \textit{requirements.txt}, the documentation \textit{README.md}, and the license \textit{LICENSE}.
    \item The \path{examples/petstore/} directory contains the Petstore OpenAPI specification, recorded HTTP request/response logs, and expected outputs.
    \item The \path{screenshots/} directory contains views of the HTML report.
\end{itemize}

To reuse RESTCov with another REST API, users provide an OpenAPI specification, a request/response log directory, and an output directory.
Dependencies and execution instructions are provided in the \texttt{README.md} file.

\section{Limitations}
Several limitations apply to RESTCov's results.
First, the demonstration uses a single API (Petstore)~\cite{swaggerapi2026swaggerpetstoreapi} for illustrative clarity.
While RESTCov has also been used across five production APIs in a comparative study~\cite{besjes2025agenticllmsrest} and an industrial study~\cite{bardakci2026promptbasedrestapi}, this paper does not present a multi-API evaluation.
Second, RESTCov reports whether documented specification elements were observed in traffic; it does not verify that endpoint behavior was correctly exercised, and its "coverage" measure should be interpreted accordingly (Section~\ref{subsec:purpose}).
Third, the current version stores each HTTP interaction as paired request/response text files, which is simple to inspect but may become cumbersome for large-scale log analysis.
Supporting consolidated log formats, such as HAR or JSONL, is left as future work rather than a current requirement.

\section{Conclusion}
RESTCov is a lightweight tool for black-box coverage analysis of REST APIs.
Given an OpenAPI specification and observed HTTP traffic, it maps executed requests back to documented API elements and reports coverage across several structural dimensions.

The JSON output supports automated analysis, while the HTML report helps inspect coverage gaps and unmatched requests.
Together, these outputs support test-suite assessment without requiring access to the service's implementation.

\balance
\bibliographystyle{IEEEtran}
\bibliography{bardakci2026SANER}

\end{document}